\documentstyle[12pt,psfig,epsfig]{article}
\newcommand{\AmS}{{\protect\the\textfont2
  A\kern-.1667em\lower.5ex\hbox{M}\kern-.125emS}}

\newcommand\numu{{\nu_\mu}}
\newcommand\anumu{\bar\nu_\mu}
\newcommand\nue{{\nu_e}}
\newcommand\anue{\bar\nu_e}
\def\dm2{\Delta m^2}
\def\sq2{sin^2(2\Theta)}
\def\nubar{\overline {\nu} }

\begin{document}

\title{Uncertainties on Atmospheric Neutrino Flux Calculations}

\author{G.  Battistoni} 

\maketitle

\begin{center}

{\small {\it I.N.F.N., Sezione di Milano, via Celoria 16, I-20133 Milano, Italy}} \\

\end{center}

\begin{abstract}
The strong evidence of new physics coming from atmospheric              
neutrino experiments has motivated  a  series of  critical
studies to test the robustness of  the  available  flux  calculations.     
In view of a  more precise determination of
the  parameters  of  new physics, new and more
refined flux calculations are in progress.
Here we review the most important sources of theoretical
uncertainties which affect these computations, and the
attempts currently under way to improve them.
\end{abstract}\vspace{0.2cm} 

\section{Introduction}
\label{sec:introduction}
The evidence for
new neutrino physics beyond the standard model, as is emerging
from the results of Super--Kamiokande\cite{sk} and other
atmospheric neutrino experiments\cite{macro,soudan2}, is now considered
robust. 
Soon after the first announcement of Super--Kamiokande in 1998, many
efforts have been devoted to the examinations of theoretical uncertainties
in the knowledge of atmospheric neutrino fluxes. None of these sources
of uncertainty resulted so critical to vanish the crucial qualitative
feature of atmospheric neutrinos: the up--down symmetry of fluxes in absence of
oscillations (or other possible mechanisms invoked to explain the observed
``anomaly'').
Even in absence of oscillations, there exist recognized violations of this absolute symmetry, as  
those due to the geomagnetic cutoff of primary cosmic rays, but they 
can be treated as additional corrections.  These perturbations are found to
be significant  mostly in the Sub-GeV region.
A new phase of the experimentation on atmospheric $\nu$'s has started, 
with the primary goal to improve and constraint as much as possible the
parameters of the proposed new physics (essentially $sin^2 2\Theta_{atm}$
and $\Delta m^2_{atm}$ for the 2--family oscillation scenario).
This is not only a goal for the existing experiments, but also the 
motivation for the proposal of new generation high precision detectors,
such as ICARUS\cite{icarus}.
For this purpose, the theoretical error has to be reduced as much
as possible and new refined calculations are necessary. As a first step
we need to improve our quantitative understanding of all the factors
which affect fundamental quantities like symmetry, flavor ratio, the
absolute flux value, the spectral 
index and the details of angular distributions.
In the following sections we intend to review
the major sources of uncertainties, with some emphasis on the
hadronic interaction sector. The work is still in progress, as
outlined in the conclusions.

\section{The status of present flux calculations}
\label{sec:present}
The present relevant experiments are making reference, for their
analyses, mainly to the neutrino flux calculations from Honda et
al. (HKKM)\cite{honda} and the Bartol group\cite{bartol}. These works 
have been recognized as the most accurate and their authors introduced
for the first time important ingredients in the simulation.
For instance the back--tracing technique for the
evaluation of geomagnetic cutoff\cite{honda} 
and the effects of muon polarization\cite{bartol}.
These calculations have in common the 1-dimensional calculation approach, 
in which all secondary particles in the showers, neutrinos included, are
considered collinear with the primary cosmic rays. This has been found
to be a non correct approximation, at least in principle. 
A new set of calculation, based on the full FLUKA simulation
code\cite{fluka}, has been recently presented\cite{3d}. There it has been
realized 
that a correct 3-dimensional approach in the earth's spherical
geometry leads to different results for the angular distribution at low energy.
A more didactic explanation of this is given in ref.\cite{lipari-geo}.
However, the new FLUKA calculation was essentially motivated by the
emerging need of a more accurate description of particle production in 
hadron and nuclear interactions. 
In fact, there are reasons to consider more critically the standard
references. For instance, it has been noticed how they 
obtain very close final results for the $\nu$-fluxes, although starting
from different particle production models and different
primary spectra. 
Another calculation appeared with a reference to FLUKA\cite{walt}, but
there the authors made use of one the hadronic interfaces 
extracted from the GEANT package v.3.21 called FLUKA. 
Actually, that package is  
only a limited and obsolete part of the hadronic model contained
in the real FLUKA code used for the present work.
It gives results which can well be different and less reliable
when compared with experimental data with respect to FLUKA.
Significant differences exist at all energies, 
but they are particularly striking
for hadron energies below a few GeV.

\section{Sources of uncertainty}
\label{sec:uncertainty}
Attempting a review of all possible sources of systematic
uncertainties in flux calculations, the following items must be
considered: primary spectra (fluxes, nuclear component, isotropy and its
breaking), geomagnetic description, atmosphere models, the geometry of
calculations, other minor details in the modeling (detector altitude, mountain
profiles, etc. ) and particle production in hadronic interactions.
In the recent past there have been other discussion (at least in part) of
these topics. 
Beyond the already quoted references \cite{3d,lipari-geo}, 
important discussions are also in 
\cite{lipari-ven99,lipari-ew,bartol-had,gaisser-comp}.

\subsection{Primary Spectrum}
\label{sec:uncertainty.pri}
\begin{figure}[tb]
\begin{center}
\mbox{\epsfysize=100mm 
      \epsffile{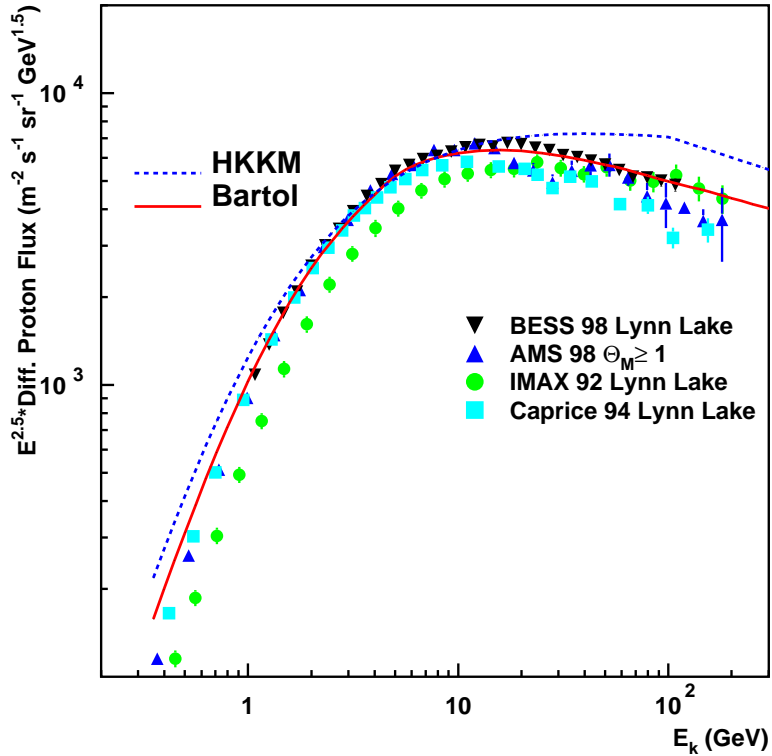}}
\end{center}
\caption{\em  Review of some of the most recent data on primary
protons. The continuous line 
represent the input model adopted in \protect\cite{honda} and
\protect\cite{bartol}. The last one was also used in
\protect\cite{3d,flukanu}. \label{fig:pr_spec}}  
\end{figure}

On of the most relevant achievements in the measurement of primary
cosmic ray spectra is the fact that BESS\cite{bess} and AMS\cite{ams}
particularly succeeded in
producing results in very good agreement one to the other, in particular
for the proton component. The scientific
community has the attitude of considering these last results as the
most reliable and therefore the uncertainty of the primary flux value is
probably smaller with respect to the estimates of few years ago, although
one should not forget other different data sets, like those of
CAPRICE\cite{caprice} until the topic is definitively settled down.
A summary of some of the recent proton measurements is shown in
Fig. \ref{fig:pr_spec}, 
together with 
the lines showing the solar minimum fits used in \cite{honda} and
\cite{bartol}. 
It has to be noticed how the input primary spectrum used by Bartol (and
later by FLUKA) is in very good
agreement with BESS data, while HKKM made use of a parameterization 
(based on the old compilation of ref.\cite{webber}) which
has a significantly higher normalization above 20$\div$30 GeV. 
It is therefore natural to ask how the
eventual result from the HKKM calculation would change if they used the
same input spectrum as Bartol. This is one of the
reasons why it is important to analyze in depth the relevance of the 
particle production model. 

As far as the Helium component is concerned, the latest AMS and BESS
measurements are now converging\cite{bess2,ams2}.
The heavier nuclei have less relevance for
low energy neutrinos; in any case, further results from AMS will hopefully
clarify the picture. 

The arguments exposed here are relevant for the energy range contributing
to contained events in Super--Kamiokande. In case of higher energy
neutrinos, like those measured through the detection of up-going muons
in MACRO and Super--Kamiokande, the uncertainties on the relevant
energy of the spectrum (up to tens of TeV) are still large (up to 20\%).

\subsection{Geomagnetic description}
\label{sec:uncertainty.mag}
The effect of
geomagnetic field is recognized as the most important source of up--down
symmetry breaking in the atmospheric neutrino flux. 
At present, the confidence in the accuracy of IGRF models is rather
strong\cite{igrf}, 
and the technique of anti-proton back--tracing
is now accepted as the standard procedure to be adopted to evaluate the correct
cutoff for primary cosmic rays arriving to the earth. 
Solar modulation has to be considered as well. Algorithm relating primary
flux to data from neutron monitors exist, and at present their quality is
considered satisfactory.
Recently, there has been
discussion about two items:
i) the role of recirculating sub-cutoff particles
as pointed out by AMS data, and ii) the anysotropy (far from earth) related to
solar wind effects at the GeV scale.
In both cases it can be demonstrated (see \cite{lipari-nu2000}) that both 
phenomena are of small relevance, since at most they affect neutrino rates
by less than 1\%.

\subsection{Geometry of calculations}
\label{sec:uncertainty.geo}
As already reported in the introduction, one of the most interesting
outcomes in last two years is the realization of the importance
of 3--Dimensional computations in a spherical geometry\cite{3d,lipari-geo}.
In summary, the net eventual result of the correct
geometrical 
description of neutrino production around the earth is a modification of
both angular distribution and of normalization in the Sub-GeV region (where.
$<\theta_{\nu - p}>$ is significant), with respect to the collinear
approximation. 
The impact of this on the determination of oscillation parameters is still
under study. In fact, a final reliable 3--Dimensional calculation of
atmospheric neutrino flux is still missing. The computations by the FLUKA
group, 
reported in \cite{flukanu}, must still be considered as preliminary, since
bending of charged particles in the geomagnetic fiels has not yet been
introduced. As discussed in \cite{lipari-ew}, this effect should play a non
negligible role. However, a detailed calculation on the whole earth sphere
taking into account also the whole B--field map introduces
technical complexities in the computation, since
spherical symmetry is lost. In fact, in absence of B--field, any point on
the earth's 
surface is equivalent to another, and this allows to make use all generated
events even for a specifc detector location, provided that a proper rotation
of trajectory parameters to the geografical coordinates of interest is
performed. 
This problem has not
yet been solved completely. In \cite{lipari-ew} a simplified solution was
proposed for the first studies, while the FLUKA group is designing a new
dedicated simulation in which specific weighting algorithms have to be
introduced. 

\subsection{Atmosphere Description and other details}
\label{sec:uncertainty.atm}
It is practically impossible to introduce a realistic description of the
atmosphere all around the earth, valid at all altitudes and for all weather
conditions. This remains an irreducible source of systematics, although
probably small. The fact that neutrino experiments last a considerable time and
detect neutrinos produced all over the earth positions, gives some
confidence on the essential validity of average atmosphere models.
A comparison between the performance of different codes, carried on by the
Bartol and FLUKA groups, is showing that having considered, or not,
ingredients in particle transport like energy loss fluctuation, multiple
scattering, etc., may affect the final results, in normalization and
flavor ratio, at the level of percent.
Other important inputs in the simulation for a specific detector site 
concern the introduction of detector altitude and possible rock overburden.
This last element, for instance, can introduce a difference in the
$\nu_e/\nu_\mu$ ratio, again at the level of some percent.

\subsection{Particle Production in Hadronic Interactions}
\label{sec:uncertainty.had}
\begin{figure}[tb]
\begin{center}
\mbox{\epsfysize=100mm 
      \epsffile{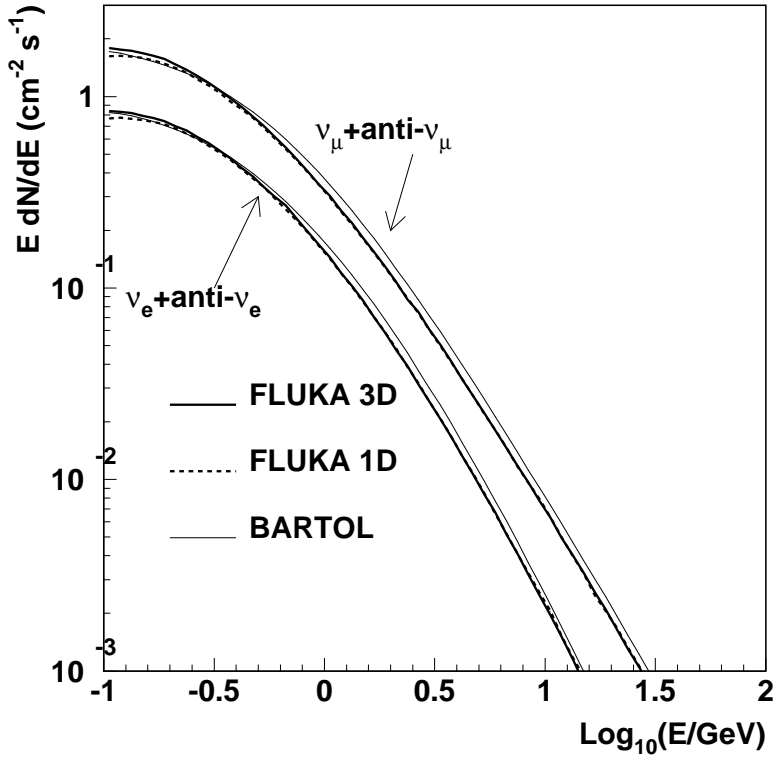}}
\end{center}
\caption{\em  Comparison of angle integrated neutrino fluxes for the
Super--Kamiokande site between Bartol and FLUKA (both 1D and 3D)
\label{fig:fluxes}} 
\end{figure}

For given shower model and primary spectrum, the most important source of
theoretical uncertainty comes from the hadronic interaction model. Since
QCD does not allow to compute the bulk properties of particle production
in the non perturbative regime, the available models are based either on 
phenomenological models, possibly inspired by partonic concepts,
constrained by accelerator data, or directly on the 
parameterizations of these experimental results.
As an example, the Bartol and FLUKA groups have used the same input primary
spectrum and, with different models, have obtained different fluxes.
The current comparison of angle integrated fluxes is reported (for the
Super--Kamiokande 
site) in Fig.\ref{fig:fluxes}, where also the 
differences between the 1D and 3D approach are shown for FLUKA.

The two groups have started a comparison of their hadronic models (FLUKA
and TARGET), both in single interaction and within the same shower code, in
order to understand the impact of different choices.
For the time being, this comparison is limited to an energy region
useful for contained and partially contained events in Super--Kamiokande.
The two codes are constructed in very different ways, and the net final
result is that there is a $\sim$20\% asymptotic
difference in the neutrino flux normalization from the two models.
In reality the difference is energy dependent: it is larger and reversed
at low energy.
The resulting spectral index is somewhat different, FLUKA being a little
harder than the Bartol one.
A comprehensive discussion of the matter should requires a dedicated paper,
and here we can only summarize a few of the crucial conclusions.

\begin{figure}[tb]
\begin{center}
\mbox{\epsfysize=100mm 
      \epsffile{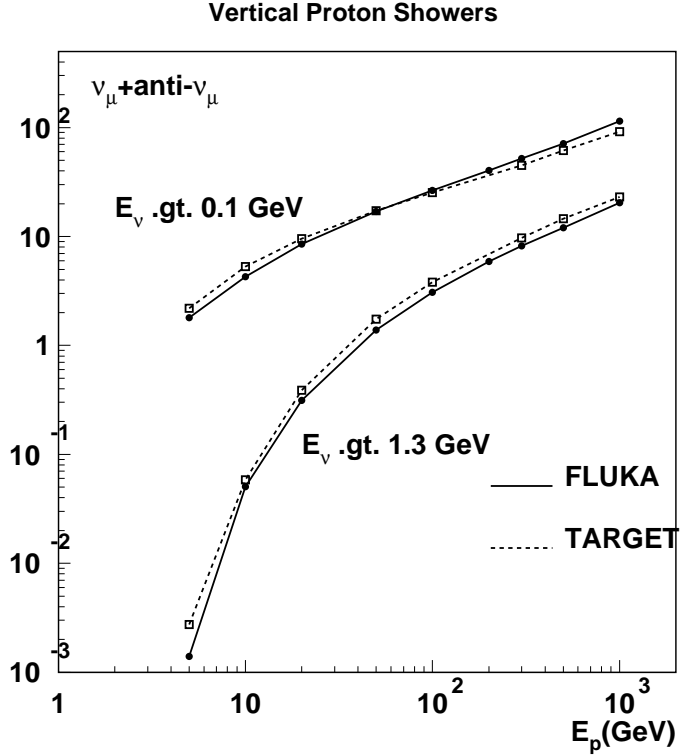}}
\end{center}
\caption{\em Average no. of atmospheric muon neutrinos for primary vertical
protons as a function of energy for the Bartol and FLUKA
models. The probability of neutrino interaction has not been considered in
this plot.\label{fig:yield}} 
\end{figure}

As far as neutrinos up to few tens of GeV are concerned, pion production in
nucleon--Nucleus interaction is the process that mostly contributes to the
yield. For known kinematical reasons, Kaon production becomes relevant at
higher energy, for instance in 
the region from which
upgoing through-muons as detected by MACRO and Super--Kamiokande are produced.
At first order, we are interested in both total multiplicity and in the
shape of the energy fraction distribution of these secondary particles, or
of Feynman--$X$ and similar longitudinal variables.
Unfortunately, although there exist reports on the total
charged multiplicity, mostly obtained in emulsion
experiments\cite{fred}, there are
not enough data on the $X$ distribution of pions in the
interactions with light nuclei. At present the most valuable data set are
those of ref. \cite{eichten} and \cite{abbot}, covering different region of
phase space. 
A direct comparison of these
data with the
predictions from FLUKA and TARGET shows that, very probably, the Bartol model
produces too many pions at low $X$.
 This difference
directly
reflect in the neutrino yield. The average number of muon neutrinos for
vertical proton showers, as a function of primary energy, is shown in
Fig.\ref{fig:yield} for the two models. It must be remembered that 
from the experimental point of view, event rates should be considered, that
is after the convolution with neutrino interaction cross sections.

Although the present data do not allow to give a reliable estimate of the
overall systematic error associated to hadronic interactions, in the
author's opinion 
the existing difference between FLUKA and TARGET ($\sim$ 20\%) should not be 
considered as 
a measurement of the real error.
The different
capability of TARGET and FLUKA model to 
reproduce accelerator data suggests that 
the actual theoretical uncertainty is likely to be definitively smaller 
than 20\%.

\begin{figure}[tb]
\begin{center}
\mbox{\epsfysize=100mm 
      \epsffile{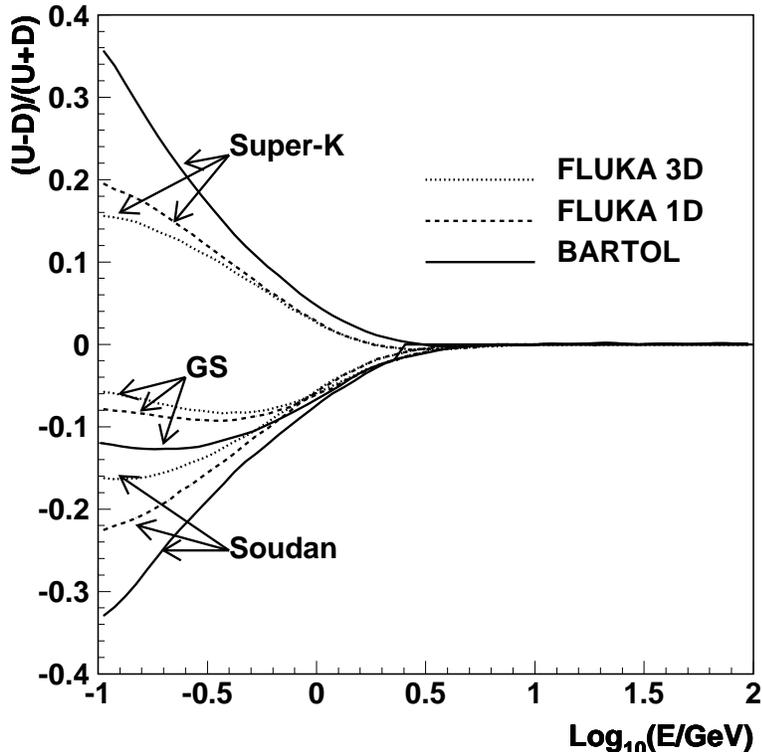}}
\end{center}
\vskip -1. cm
\caption{\em Up--Down asymmetry of $\numu$ fluxes, in absence of
oscillations, at 3 different geomagnetic latitudes as calculated with the
Bartol and FLUKA (1D and 3D) models. The probability of neutrino
interaction has not been considered in this plot. The differences in
asymmetry of event rates, after the convolution with neutrino interaction
cross sections, are smaller.\label{fig:asym}} 
\end{figure}

A preliminary comparison, at the level of single interaction features, with
the models adopted in HKKM calculations, gives indications that, if they
had used the same primary spectrum of Bartol and FLUKA, also in their case the
normalization would have been lower.
The yield difference between FLUKA and TARGET is larger at lower nucleon
energy. This has some direct consequence in the prediction of up-down
symmetry of fluxes in different geographical sites. For instance, at
Super--Kamiokande, where the cutoff from the above direction is rather high
(around 10 GeV), the yield enhancement of TARGET is ineffective, and the
FLUKA/TARGET ratio in the Sub-GeV region is reversed (see also Fig.\ref{fig:fluxes})
with
respect to the situation of Soudan, which is instead a low--cutoff site.
Therefore, since the analysis of Soudan data are based upon the Bartol
predictions, they should expect a lower asymmetry in the Sub-GeV region:
see Fig.\ref{fig:asym}

\begin{figure}[phtb]
\begin{center}
\begin{tabular}{cc}
\mbox{\epsfysize=75mm 
      \epsffile{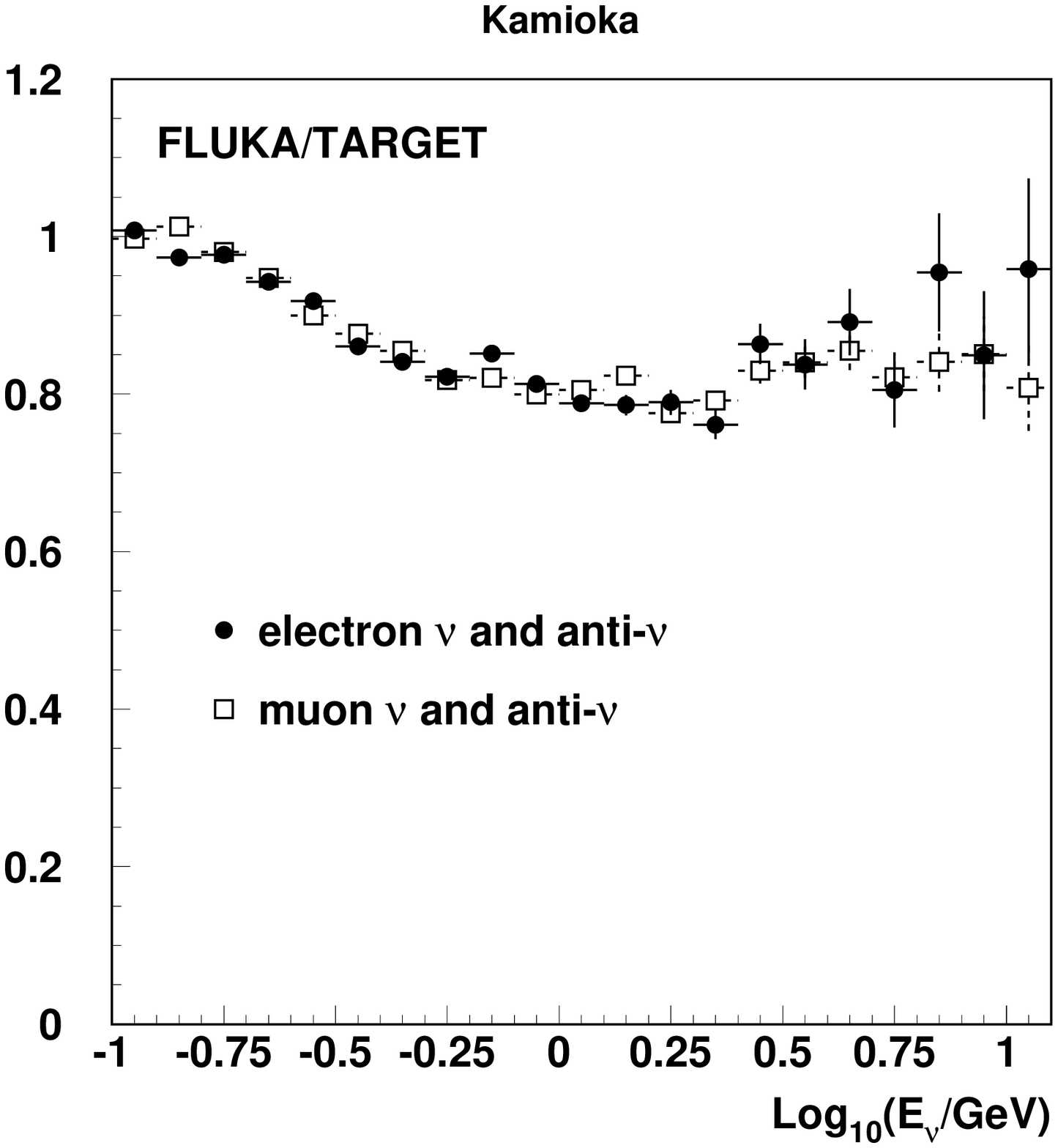}} &
\mbox{\epsfysize=75mm 
      \epsffile{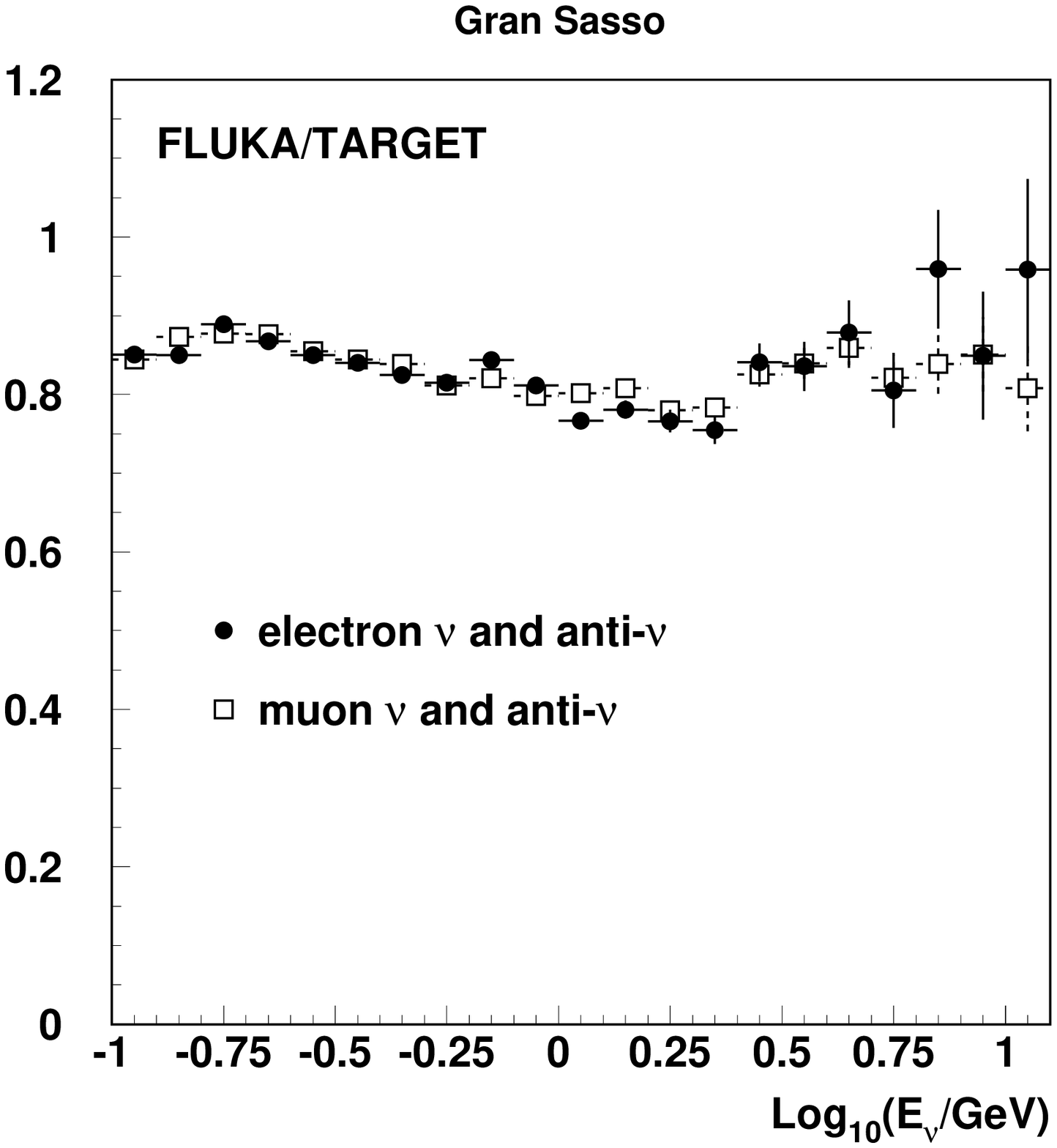}} \\
\multicolumn{2}{c}{
\mbox{\epsfysize=75mm 
      \epsffile{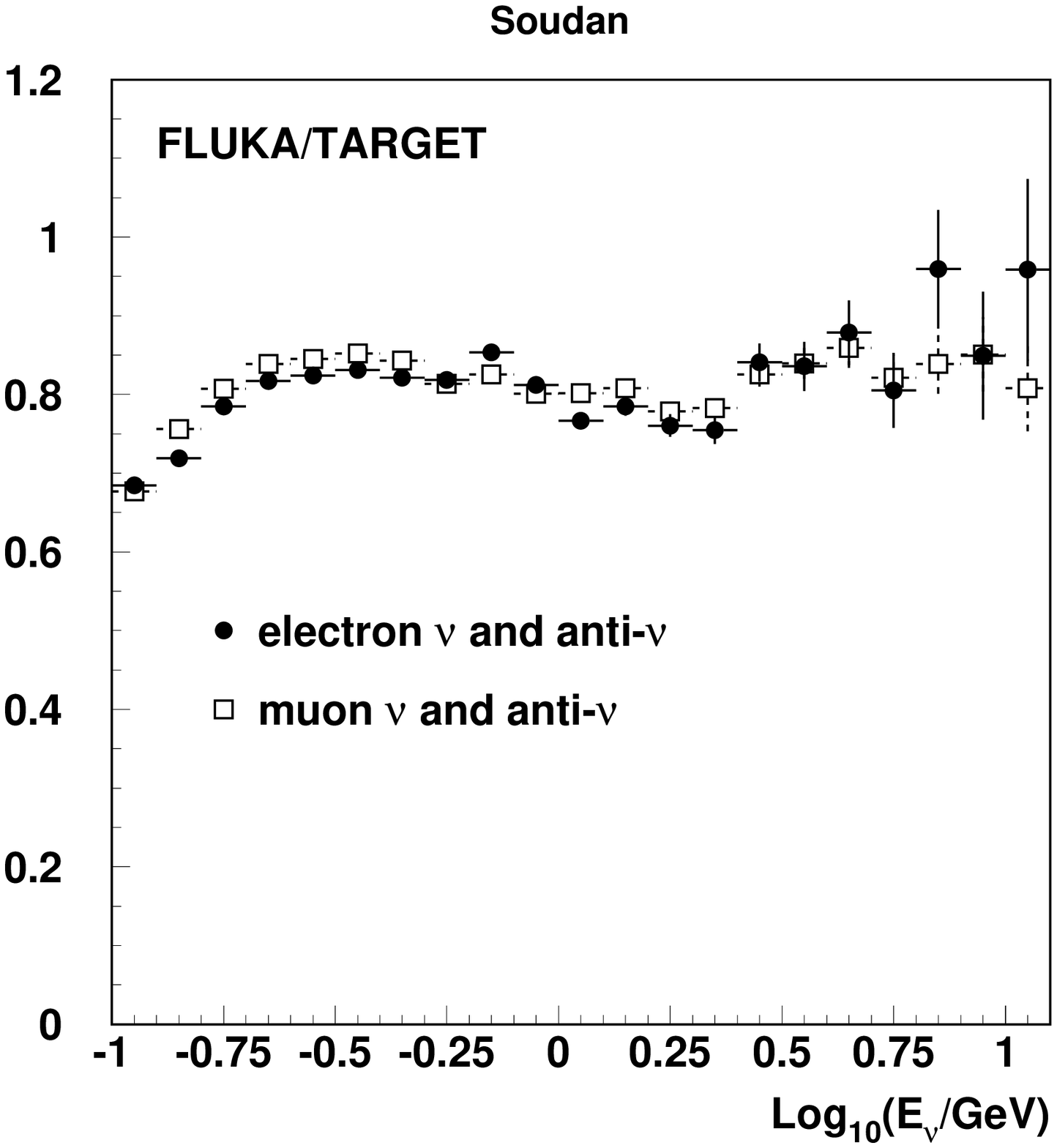}} } \\
\end{tabular}
\end{center}
\vskip -1. cm
\caption{\em Ratio of FLUKA (1D) to TARGET (inside FLUKA 1D shower code) calculated $\numu$ and
$\nue$ fluxes, in absence of
oscillations, as a function of neutrino energy, for 3 different geomagnetic
latitudes.
\label{fig:asym2}} 
\end{figure}

Another way of looking at this is given in Fig.\ref{fig:asym2}, where
the ratio of FLUKA (1D) to TARGET fluxes are shown as a function of energy for
the different laboratories.

The two interaction models also differ in the $\pi^+/\pi^-$ ratio, which
affects the $\nu/\nubar$ ratio. Due to the different interaction cross
section this is a significant parameter. Considering the proper weighting
with the $X$ distribution, FLUKA has a larger $\pi^+/\pi^-$ ratio by an
amount which is around 7\% for protons at 10 GeV. For increasing energy
the difference becomes smaller, as expected. FLUKA is able to
satisfactorily reproduce the charge ratio measured by many
experiments\cite{ferrari}. 

We have found instead that the $(\nue + \anue)/(\numu + \anumu)$
ratio has not a  relevant dependence on the hadronic interaction model.

Oscillation analysis of neutrino events requires the knowledge of
production height, which depends on the longitudinal development of
showers. This, on turn, depends on the point of first interaction,
determined by total 
inelastic cross sections, and by the following development driven
by the energy fraction carried away by leading nucleons in each
interaction. Total cross sections are eventually determined by
nucleon--nucleon cross sections, which are well constrained by existing
data compilation, to which all groups make strong reference. Of course,
the energy fraction taken by pions 
(including $\pi_0$, feeding the e.m. component of showers) is not
independent from that taken by nucleons. FLUKA and TARGET have different
elasticities and therefore give rise to some differences in the
longitudinal development of cascades. Again, the comparison with existing
data reinforces some confidence on the FLUKA model.
More
complete experimental data on particle production on light nuclei would be
fundamental to minimize the theoretical uncertainties and constraint the
existing models. 
In order to be significant for this purpose, 
a new experiment must explore
a range of beam energies from few GeV up to at least
30 $\div$ 50 GeV, with targets of different atomic number,
from Be up to at least Al, in order to study
the dependence on the number of
elementary collisions (which
in the Glauber approach scales as $A^{1/3}$\cite{glauber}). Secondary
particles must be 
measured in a wide solid angle to cover as much as possible the available phase
space.
For these reasons the
scientific community welcomes the HARP experiment\cite{harp}, proposed to
perform a 
dedicated study on these subjects.

\section{Conclusions}
\label{sec:conclusions}
The current understanding of the sources of uncertainty in
atmospheric neutrino flux calculations is still improving.
A better confidence on the primary cosmic ray spectra is the first
necessary condition. After that, the
most important source of uncertainty is that due to the particle production
model. 
There is discussion upon which kind of experiment or study can help in 
achieving better constraining of simulations. 
Data on muon fluxes in atmosphere can help, and they are a useful benchmark
tool. However, the connection between muon and neutrino yield at different
altitudes and energy is still a rather indirect process, and it is not yet
clear if muon balloon experiments can 
guarantee a level of systematics below $\sim$10\%. In the author's opinion,
the impact of new
data from accelerators could be more relevant on model building, 
provided that in this case systematics is kept under better control.
In this will be the case,
there  are reason to believe that, although the theoretical error
cannot be erased, it could be reduced to the
10\% level or even less. 
However, one of the most serious problem could stay not only in the flux
calculations, 
but also in the
knowledge of absolute neutrino cross sections with nuclei, expecially
 at low energy for quasi--elastic scattering and resonance production.
As a matter of fact, the experimentalists are more interested in the
eventual event rates than in the flux itself. 
There is the
serious risk  that there will be no experimental answer to clarify this
aspect.

Other improvements in these computations are needed,
but they will require more and more efforts. As an example, the FLUKA group
is planning to make use of further developments in particle production
models. In particular we are aiming to study the effect of nuclear
projectiles instead 
of recurring to the usual incoherent addition of nucleons (superposition
model) adopted so far also in the other standard references.
The impact of more precise calculations on the measurement of
the parameters of ``new physics'' is still to be understood in
detail. In our opinion, this can be reliably done only introducing the correct simulations of the
actual experiments, since detector sources of systematics and resolutions
are likely to be of an importance comparable to that of theoretical
factors. 

\section{Acknowledgments}
\label{sec:acknowledgments}
The author is indebted to the other authors of the FLUKA--neutrino
calculation, A. Ferrari, T. Montaruli, and. P.R. Sala, for the help received
in preparing this review.
This work has been made possible thanks to 
the collaboration with the Bartol group: R. Engel, T.K. Gaisser, P. Lipari and
T. Stanev. They have provided the TARGET model and a lot of essential data
and suggestions. 

{ }
\end{document}